\documentclass[a4paper,showkeys,floatfix,aps,pre,reprint,groupedaddress]{revtex4-1}
\usepackage{amsmath,amssymb}
\usepackage{graphics,graphicx}
\usepackage{dcolumn,bm}
\usepackage{psfrag}
\usepackage{soul}

\begin{document}

\title{Predictability and Strength of a Heterogeneous System : \\ The Role of System Size and Disorder}

\author{Subhadeep Roy}
\email{sroy@eri.u-tokyo.ac.jp}
\affiliation{Earthquake Research Institute, University of Tokyo, 1-1-1 Yayoi, Bunkyo, 113-0032 Tokyo, Japan.}

\date{\today}

\begin{abstract}
\noindent 
In this work I have studied the effect of disorder ($\delta$) and system size ($L$) in fiber bundle model with a certain range $R$ of stress redistribution. The strength of the bundle as well as the failure abruptness is observed with varying disorder, stress release range and system sizes. With a local stress concentration, the strength of the bundle is observed to decrease with system size. The behavior of such decrement changes drastically as disorder strength is tuned. At moderate disorder, $\sigma_c$ scales with system size as: $\sigma_c \sim 1/\log L$. In low disorder, where the brittle response is highly expected, the strength decreases in a scale free manner ($\sigma_c \sim 1/L$). With increasing $L$ and $R$ the model approaches thermodynamic  limit and the mean field limit respectively. A detail study shows different limits of the model and the corresponding modes of failure on the plane of above mentioned parameters ($\delta$, $L$ and $R$).          
\end{abstract}


\maketitle

\section{Introduction}
Fluctuation in strength ($\sigma_c$) of materials is studied extensively in the field of engineering and material science in the last decade. The fluctuation in $\sigma_c$ is observed in real system with length ($L$) of the material under observation \cite{Bazant,Weibull,Danzer,Basua,Danzer1}. Such fluctuation depends on the mode of failure, which in tern is governed by two crucial factors : (i) the concentration of defects or disorder in the material and (ii) the stress field morphology in presence of the defects and its evolution as the fracture develops in the material. It is known that defects like micro cracks, dislocations or grain boundaries vastly reduce the strength of a material, as large stresses can develop at the sharp edges of the micro-cracks and the motion of dislocation helps the crystal planes to slip easily on each other \cite{Dieter}. Griffith suggested \cite{Lawn} that the typical stress at which a sharp micro-crack of length $l$ will become unstable and grow to break the material decreases as $1/\sqrt(l)$ \cite{Bazant}. This led to the weakest link of a chain concept \cite{Tanaka,McClintock,Ray}, according to which the fracture in presence of many defects is determined primarily by the most vulnerable defect and that is how the idea of extreme statistics is applied to fracture. In presence of many defects, the failure process is not guided by a single micro-crack like the Griffith's theory, instead the process is determined by the interection of all defects. In such a case the material indeed shows a large scatter in fracture strength, the distribution of which can be represented by the long-tailed Weibull distribution. The distribution is a telltale signature of the underlying extreme events that controls triggering of fracture. The distribution suggests power law fall of the fracture strength with system size which is readily seen in experiments \cite{Bazant1,Bardenhagen,Carpinteri}. Apart from long-tailed Weibull distribution, experiments in different materials also show Gumbell distribution. Here in-stead of power law, $\sigma_c$ falls logarithmically ($\sim 1/\log L$) \cite{Duxbury}. Whether the fracture strength should follow Weibull, Gumbell or some other distribution is a matter of debate for long time. For example, it was claimed in ref \cite{Born} that a Gumbell distribution (also known as Duxbury-Leath distributions) fits more appropriately with the strength distribution for a highly porous ceramics. In contradictory, ref \cite{Vliet} suggests a much better fit with Weibull distribution in case of concrete and sandstone under uniaxial tension. One early paper has already discussed about the fact which distribution, Weibull or normal, is appropriate for nominal stress in real specimens \cite{Lu}. In this paper we have dealt with a statistical model of fracture, namely the fiber bundle model, and studied how the system size effect on strength and failure abruptness varies with disorder present in the model. Though a previous study \cite{Pradhan} in fiber bundle model has investigated the behavior of nominal stress at low disorder, a complete study of both strength and failure abruptness is still absent.

%%%%%%%%%%%%%%%%%%%%%%%%%%%%%%%%%%%%%%%%%%%%%%%%%%%%%%%%%%%%%%%%%%%%%%%%%%%%%%%%%

\section{Description of the Model}
Fiber bundle model \cite{Duxbury1,RMP} consists of vertical fibers (mostly considered as Hookean) attached between two parallel bars. The bars are pulled apart with a force $F$, applying a stress $F/L$ on $L$ fibers ($L$ is also considered as the size of the system). A periodic boundary codition is applied perpendicular to the direction of applied force. Disorder in the bundle is introduced as the fluctuation of strengths among individual fibers. Each fiber sustains a stress up to a threshold, chosen randomly from a  distribution, beyond which it breaks irreversibly. Width of such distribution is the amount of disorder in the model. Once a fiber is broken, the extra stress of the broken fiber is distributed equally among the rest of the model. We assume a range $R$, up to which the stress of the broken fiber is redistributed. $R$ is basically the number of surviving nearest neighbors on the other sides of the broken one (in total $2R$ fiber will experience the extra stress). After such redistribution, the local increment in stress profile might cause failure of the neighbors and the process may continue till all fibers fail. Otherwise, the system can attain a equilibrium with few broken fibers, when a further increment of external stress is required to make the process evolve. With increasing applied stress, the model breaks in avalanches (burst of fibers). The stress at which this total breakdown happens is known as critical stress or strength $\sigma_c$ of the model . The fraction of unbroken bonds at this stress is the critical fraction ($n_c$) and also this is a quantitative measurement of fracture abruptness ($n_c\approx1$ is the signature of abrupt failure). A limit $R=1$ (one fiber on each side, two in total) coincides with local load sharing (LLS) scheme\cite{Phoenix,Smith,Newman,Harlow2,Harlow3,Smith2}, while with the same convention, $R \approx L/2$ is the mean field limit \cite{Pierce,Daniels} of the model.

%%%%%%%%%%%%%%%%%%%%%%%%%%%%%%%%%%%%%%%%%%%%%%%%%%%%%%%%%%%%%%%%%%%%%%%%%%%%%%%%%

\section{Numerical Results}
Numerical results are discussed here for one-dimensional bundle with system sizes ranging from $10^3$ to $5\times10^4$. An uniform threshold distribution, of mean at 0.5 and half-width $\delta$, is used for numerical simulation. The disorder is tuned numerically by tuning this half width. With varying $R$, the model is made to travel from pure localized stress concentration limit (LLS, $R=1$) to the mean field limit ($R\approx L/2$). On the other hand increasing system size leads to thermodynamic limit where only brittle like abrupt failure is observed at any disorder value (reference \cite{Moreira} and \cite{Shekhawat} have discussed such thermodynamic limit in the random resistor network).

%%%%%%%%%%%%%%%%%%%%%%%%%%%%%%%%%%%%%%%%%%%%%%%%%%%%%%%%%%%%%%%%%%%%%%%%%%%%%%%%%

\subsection{Study of failure abruptness}
\begin{figure}[ht]
\centering
\includegraphics[width=6cm, keepaspectratio]{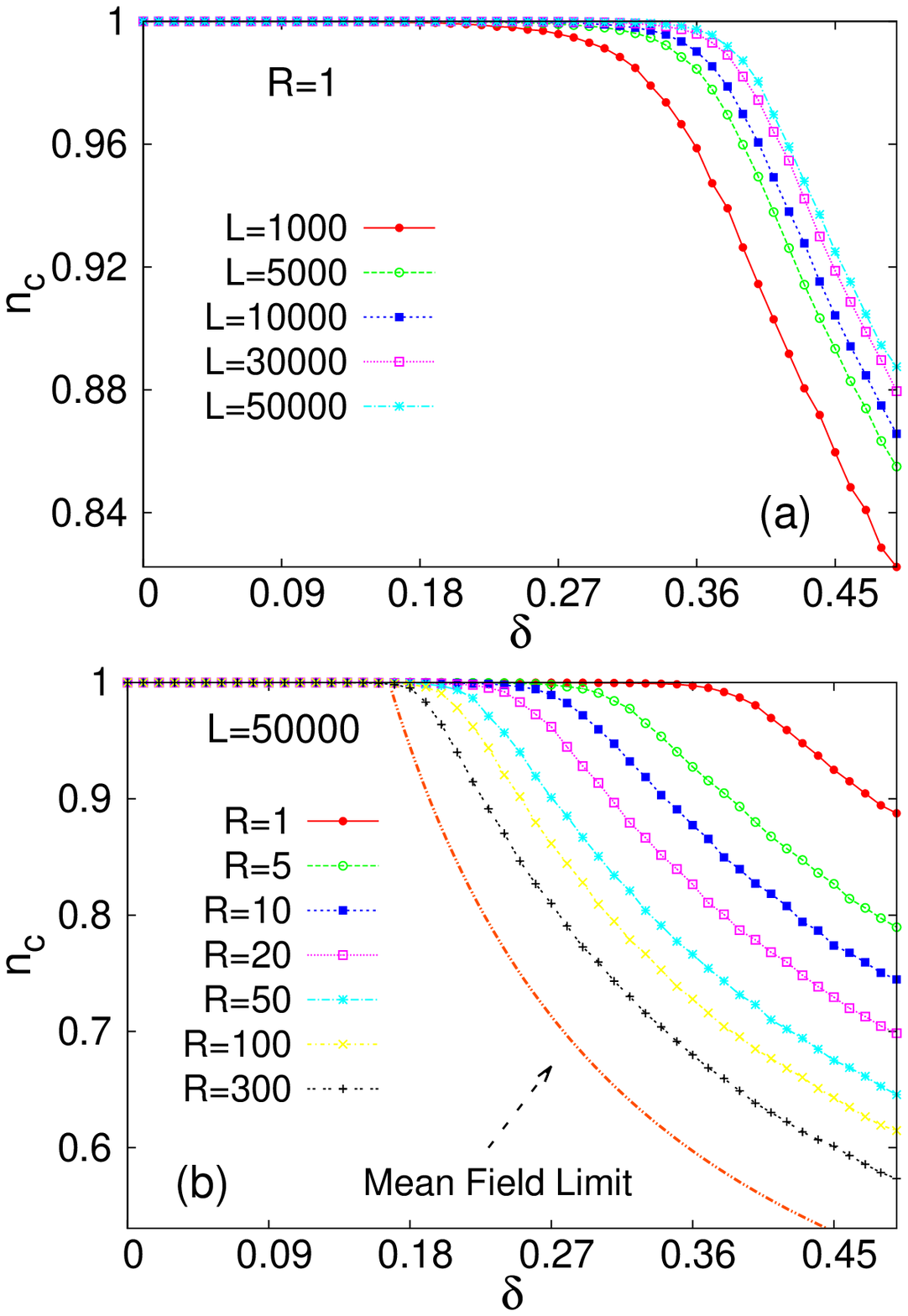}
\caption{(Color online) Study of fraction of unbroken bond with varying disorder and stress release range. \\ (a) $n_c$ v/s $\delta$ at $R=1$ and system sizes ranging from $10^3$ to $5\times10^4$. $n_c$ remains at $1$ at low disorder and decreases continuously at high disorder. At a constant $\delta$, $n_c$ increases with increasing $L$ values. \\ (b) $n_c$ v/s $\delta$ at $L=5\times10^4$ and for $1 \le R \le 300$. With increasing range value the model approaches its mean field limit.}
\label{fig:Abruptness}
\end{figure}

Figure \ref{fig:Abruptness} shows the behavior of fraction of unbroken bonds ($n_c$) corresponding to the critical point, at a constant range and different system sizes (Fig.\ref{fig:Abruptness}a) as well as at a constant system size and different stress release range values (Fig.\ref{fig:Abruptness}b).  
$n_c$ is actually a measurement of abruptness of failure process. $n_c=1$ suggests that the total model was intact just before global failure and hence suggests an abrupt brittle like failure. The failure process become quasi-brittle like non abrupt when $n_c$ deviates from 1.0. Fig.\ref{fig:Abruptness}(a) shows that the failure process is abrupt at low disorder values. The maximum $\delta$ value up to which $n_c$ remains close to 1 is known as $\delta_a$, the limit of abrupt failure. At a constant $R$, $\delta_a$ shifts to higher value when the system size is increased. In such a process, the model slowly approaches the thermodynamic limit ($L\rightarrow\infty$), where the failure process is abrupt at any disorder value. 

The variation of $n_c$ with $R$ is shown in Fig.\ref{fig:Abruptness}(b), when the system size is kept constant at $L=5\times10^4$. With increasing stress release range, the model entires the mean field limit (shown by the dotted line in Fig.\ref{fig:Abruptness}b) beyond a critical range value $R_c$ \cite{Biswas}. In the mean field limit the system size effect vanishes and we get an unique $\delta$ value $\delta_c^{mf}(=1/6)$ \cite{sroy}, below which the failure is abrupt (like brittle) and above the which the model breaks in avalanches (like quasi-brittle). 
  
\begin{figure}[ht]
\centering
\includegraphics[width=6cm, keepaspectratio]{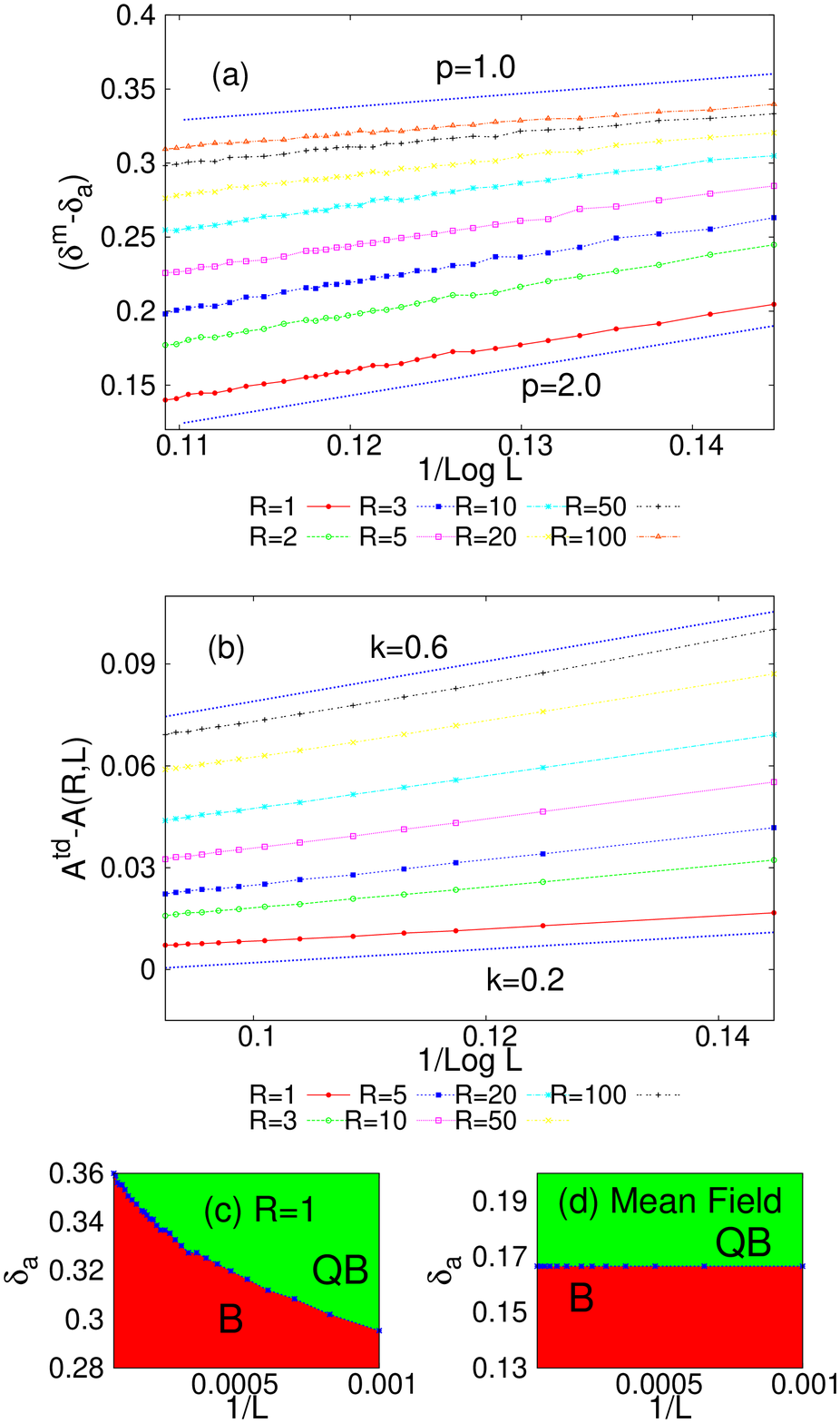}
\caption{(Color online) (a) Variation of $\delta_a$ with $L$ at different range values. $\delta_a \sim p(R)/\log{L}$, where $p(R)$ decreases with increasing $R$ values. \\ (b) Variation of $A(R,L)$ with $L$ at different $R$ values. $A(R,L)=A^{td}+k(R)/\log{L}$, where $A^{td}$ is the value of $A(R,L)$ at thermodynamic limit and $k(R)$ increases with increasing $R$ values. \\ (c)-(d) Comparison of $\delta_a$ v/s $1/L$ behavior for $R=1$ and in the mean field limit.}
\label{fig:Dectac_PhaseDiagram0}
\end{figure}
For further quantitative measurement of failure abruptness we have studied $\delta_a$ with a continuous variation of $R$ and $L$. Other than $\delta_a$, another measurement of abruptness is $A(R,L)$, the area under $n_c$ v/s $\delta$ plot at a particular range $R$ and system size $L$. $A(R,L)$ is defined as follows:
\begin{align}
A(R,L)=\displaystyle\int_{0}^{0.5}n_cd\delta=&\displaystyle\int_{0}^{\delta_a}n_c(\text{brittle})d\delta+ \nonumber \\
                                                        &\displaystyle\int_{\delta_a}^{0.5}n_c(\text{quasi brittle})d\delta \nonumber
\end{align}
where $\delta_a$ is a function of both $L$ and $R$. We already know that $n_c(\text{brittle})=1$, while $n_c(\text{quasi brittle})$  has an analytical expression in the mean field limit (for uniform distribution) \cite{sroy}. In the thermodynamic limit, $n_c=1$ throughout the region $0\le\delta\le0.5$.
\begin{itemize}
\item The area in thermodynamic limit will be,
\begin{align}
A^{td}=\displaystyle\int_{0}^{0.5}(1)d\delta=0.5 \nonumber
\end{align}
\item On the other hand, in the mean field limit,
\begin{align}
A^{mf}&=\displaystyle\int_{0}^{\delta_c^{mf}}(1)d\delta+\displaystyle\int_{\delta_c^{mf}}^{0.5}\displaystyle\frac{\delta}{2}\left(1+\displaystyle\frac{c-\delta}{2\delta}\right)d\delta \nonumber \\
&=0.387376 \nonumber
\end{align}
\end{itemize}    
where $c$ is the mean of the uniform threshold distribution (for numerical simulation we have chosen $c=0.5$). Figure \ref{fig:Dectac_PhaseDiagram0} shows the variation of $\delta_a$ as the model approaches the thermodynamic limit with increasing system sizes.
\begin{align}
\delta_a=\delta^m-\displaystyle\frac{p(R)}{\log{L}} \ \ \ \ \ \text{and} \nonumber \\ 
A(R,L)=A^{td}-\displaystyle\frac{k(R)}{\log{L}} 
\end{align}
In the limit $L\rightarrow\infty$, $\delta_a\rightarrow\delta^m$, the maximum value of the disorder. In case of uniform distribution $\delta^m=0.5$, when the thresholds are redistributed between 0 and 1. This suggests that only brittle like failure is observed in such limit. At a finite system size, $\delta_a$ decreases logarithmically with $L$ (Fig.\ref{fig:Dectac_PhaseDiagram0}a). Also the area $A(R,L)$ approaches $A^{td}$ in the same fashion (Fig.\ref{fig:Dectac_PhaseDiagram0}d). Fig.\ref{fig:Dectac_PhaseDiagram0}(c) and Fig.\ref{fig:Dectac_PhaseDiagram0}(d) shows $\delta_a$ v/s $L$ variation, between mean field limit and local load sharing limit ($R=1$). As already observed \cite{sroy}, in mean field limit $\delta_a$ remains constant at $\delta_c^{mf}$. Here $\delta_c^{mf}$ separates a constant amount of brittle (B, red color) and quasi-brittle (QB, green color) region, irrespective of the system size. In the region $R<R_c$ (see ref \cite{Biswas}) $\delta_a$ increases with increasing $L$ values, and we get more brittle response for a particular disorder value as we increase $L$. The origin of figure \ref{fig:Dectac_PhaseDiagram0}(c) corresponds to the thermodynamic limit where only brittle failure is observed. 

%%%%%%%%%%%%%%%%%%%%%%%%%

\subsection{Strength of the bundle : \\ Effect of system size and disorder}
Next we have studied how the strength of the bundle varies with system size as well as with the disorder width $\delta$. For such study we have chosen $R=1$, since at higher $R$ values the model approaches the mean field limit where system size effect is absent.
\begin{figure}[ht]
\centering
\includegraphics[width=6cm, keepaspectratio]{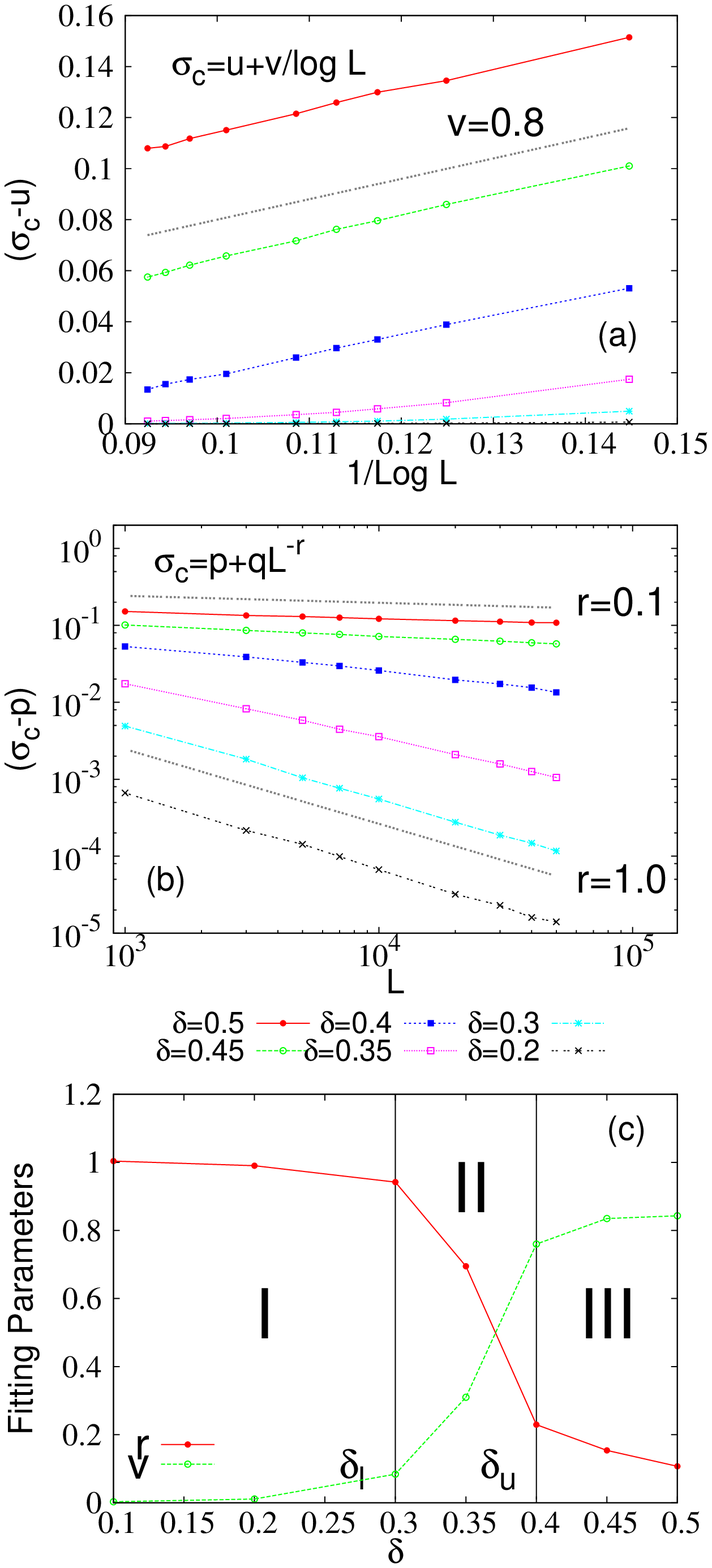}
\caption{(Color online) Study of the system size effect of $\sigma_c$ at different disorder values. \\ (a) $\sigma_c$ is fitted with $u+v/\log{L}$ at different disorder. The fitting is satisfactory at high disorder limit ($\ge0.4$). \\ (b) $\sigma_c$ is fitted with $p+qL^{-r}$ for all disorder values in between 0.1 and 0.5. \\ (c) The exponent $r$ decreases with increasing disorder. $r$ shows a sudden change within the window $\delta_l \le \delta \le \delta_u$.}
\label{fig:System_Size_Variation1}
\end{figure}

\begin{figure}[ht]
\centering
\includegraphics[width=6cm, keepaspectratio]{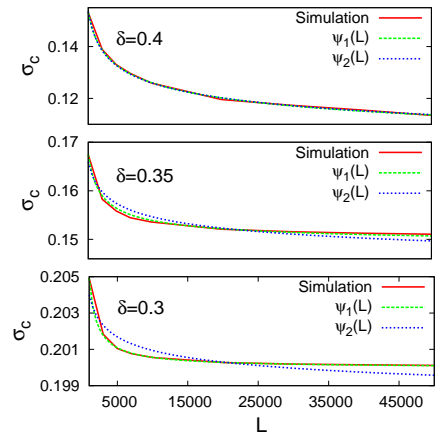}
\caption{(Color online) $\sigma_c$ as a function of $L$ is fitted with $\Psi_1(L)$ and $\Psi_2(L)$ for three different strength of disorder (0.3,0.35,0.4) within the window [$\delta_l,\delta_u$].}
\label{fig:Powerlaw_Logarithmic_Fit}
\end{figure}

So the main questions are : (i) how $\sigma_c$ varies with system sizes ? and (ii) how this variation is affected by the disorder present in the model ?  From the experimental findings, we already know about the existence of both scale free and logarithmic decrease of $\sigma_c$ with increasing system size \cite{Born,Vliet,exp2}. In the light of such experimental findings we have fitted the strength of the bundle with the following two functions: 
\begin{align}\label{Eq.2}
&\sigma_c=\Psi_1(L)=p+qL^{-r} \ \ \ \ \ \text{and} \nonumber \\
&\sigma_c=\Psi_2(L)=u+\displaystyle\frac{v}{\log{L}}
\end{align}
where the parameters $p$, $q$, $r$, $u$ and $v$ depends on the disorder $\delta$. $p$ and $u$ are the critical stress corresponding to the thermodynamic limit. For the present distribution (mean at 0.5 and half-width $\delta$), $p$ and $u$ will be given by ($0.5-\delta$). With varying disorder value, we have found an window $[\delta_l,\delta_u]$ ($l$ and $u$ stands for lower and upper limit respectively. See figure \ref{fig:System_Size_Variation1}c) where most of the changes in the behavior of $\sigma_c$ is observed. Below, we have discussed our findings in three different regions observed in the model (see figure \ref{fig:System_Size_Variation1}c). 
\begin{itemize}
\item \underline{Region I ($\delta<\delta_l$)} : Figure \ref{fig:System_Size_Variation1} shows that, below $\delta_l$, $\sigma_c$ matches much better with $\Psi_1(L)$. In this region $\sigma_c$ falls with increasing system size in a scale free manner. The exponent $r$, of such scale free decrease, has a value close to 1.0 and it remains constant throughout the region. The logarithmic fit in this region is not satisfactory at all.    
\item \underline{Region III ($\delta>\delta_u$)} : Beyond $\delta_u$, $\sigma_c$ fits quite well with both $\Psi_1(L)$ and $\Psi_2(L)$. In this region $\sigma_c$ shows both logarithmic decrease and scale free decrease with system sizes. Here the exponent of the scale free decrease is as low as 0.1. Both the behavior matches most probably because we know that $x^{\alpha}$ behaves as $\log{x}$ for low $\alpha$ values. As a result, it is not possible to distinguish between logarithmic, and scale free behavior in this region.
\item \underline{Region II ($\delta_l \le \delta \le \delta_u$)} : In the region $\delta_l \le \delta \le \delta_u$, $\sigma_c$ fits with $L^{-r}$ but the exponent $r$ decreases continuously with increasing disorder value. $r$ has a value 1.0 close to $\delta_l$ and gradually reaches 0.1 at $\delta=\delta_u$ (see figure \ref{fig:System_Size_Variation1}b). Also $\Psi_2(L)$ does not fit satisfactorily in this region. 
\end{itemize}  
The fitting of the critical stress with the functions $\Psi_1$ and $\Psi_2$, given by Eq.\ref{Eq.2}, is shown explicitly in Fig.\ref{fig:Powerlaw_Logarithmic_Fit}. I have provided the results within the window [$\delta_l,\delta_u$] to show the continuously changing behavior of the system size effect of $\sigma_c$.   
Fig.\ref{fig:Powerlaw_Logarithmic_Fit} shows that, close to the lower limit of above mentioned window ($\delta=\delta_l=0.3$), the numerical result fits well with the scale-free functional form $\Psi_1(L)$. The fitting of the results with $\Psi_2(L)$ is not satisfactory here. As we increase $\delta$ both function starts converging with each other (see the middle figure of Fig.\ref{fig:Powerlaw_Logarithmic_Fit} where $\delta=0.35$). Finally near the upper limit of the window ($\delta=\delta_u=0.4$), both the function $\Psi_1(L)$ and $\Psi_2(L)$ fits satisfactorily with the numerical results.    

%%%%%%%%%%%%%%%%%%%%%%

\subsection{Thermodynamic limit of the model}
\begin{figure}[ht]
\centering
\includegraphics[width=7.5cm, keepaspectratio]{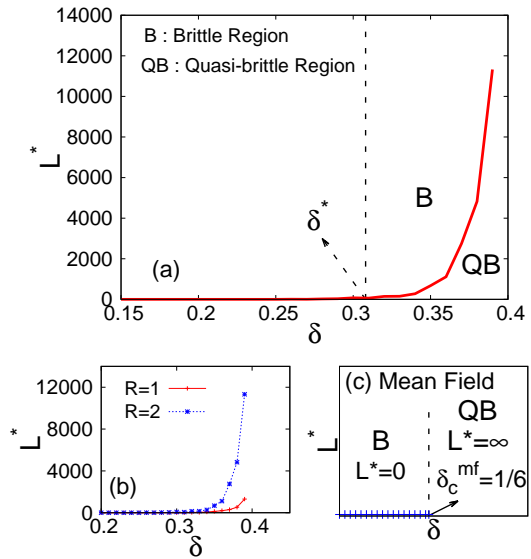}
\caption{(Color online) (a) Variation of $L^{\ast}$ with disorder $\delta$. The region $L>L^{\ast}$ is the brittle region. The region $\delta<\delta^{\ast}$ is the pure brittle region while for $\delta>\delta^{\ast}$ both brittle and quasi-brittle region exists depending on system size. \\ (b) $\delta$ vs $L^{\ast}$ for two stress release range $R=1$ and $2$. \\ (c) A schematic diagram of mean field limit where we get to distinct modes of failure, brittle and quasi-brittle, irrespective of the system size.}
\label{fig:Lmax_Delta}
\end{figure}
Due to the system size effect of failure abruptness (in the region $R<R_c$), we can not define a particular mode of failure at a certain disorder value. At any disorder we can go from non-abrupt failure to abrupt failure just by increasing the system size. As a result, unlike mean field limit, we do not have an unique disorder here that separates this two modes of failure. At a constant $\delta$, we define $L^{\ast}$ as a special system size value, above which the failure process is abrupt (like brittle materials). Below $L^{\ast}$ we get quasi-brittle response where the failure happens continuously in avalanches. Figure \ref{fig:Lmax_Delta} shows the behavior of $L^{\ast}$ with continuous variation of $\delta$. We observe a special disorder value $\delta^{\ast}$, below which the failure is always abrupt. Figure \ref{fig:Lmax_Delta}(a) shows the variation of $L^{\ast}$ with $\delta$ at a particular range $R$ (we have chosen lowest possible $R$ values for this purpose). Below $\delta^{\ast}$ we only get brittle like abrupt response even when the system size is extremely low. This is the pure brittle region. The system size effect comes into the picture beyond this $\delta^{\ast}$. For $\delta>\delta^{\ast}$, we get both brittle (B) and quasi-brittle (QB) region depending on the size of the bundle. Figure \ref{fig:Lmax_Delta}(b) shows the effect of stress release range on such system size effect above $\delta^{\ast}$. When we increase $R$, mainly two changes is being observed: (i) $\delta^{\ast}$ approaches $\delta_c^{mf}$ as expected \& (ii) For any $\delta>\delta^{\ast}$, we have to reach a relatively higher L value to enter the brittle region. We have shown the results for $R=1$ and $R=2$. As we go to higher $R$ values, the simulation becomes understandably more and more time consuming since $L^{\ast}$ will have very high values there. With such increasing $R$ values we finally reach the mean-field limit where $\delta^{\ast}\approx\delta_c^{mf}(=1/6)$. Fig.\ref{fig:Lmax_Delta}(c) shows the schematic diagram drawn from the knowledge of $L^{\ast}$ with increasing $R$ values. At mean field limit, $L^{\ast}\rightarrow 0$ for $\delta<\delta_c^{mf}$ and $L^{\ast}\rightarrow \infty$ for $\delta>\delta_c^{mf}$. This allows us to clearly demarcate between abrupt (brittle like) and non-abrupt (quasi-brittle like) failure on the other side of $\delta_c^{mf}(=1/6)$.  

%%%%%%%%%%%%%%%%%%%%%%

\subsection{Power-law stress redistribution scheme} 
Along with the uniform redistribution scheme (over a range R), a set of results with a relatively realistic version of stress redistribution is also produced. Here, instead of uniformly, the stress of a broken fiber is redistributed in a scale free manner up to a range $R$.
\begin{figure}[ht]
\centering
\includegraphics[width=6cm, keepaspectratio]{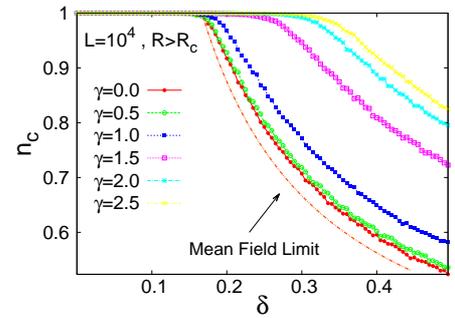}
\caption{(Color online) Variation of $n_c$ with $\delta$ for different $\gamma$ values. The stress release range is kept above $R_c$. The failure becomes more abrupt as the stress localization increases with increasing $\gamma$. On the other low $\gamma$ values lead the model towards the mean field limit.}
\label{fig:Abruptness_Plwerlaw}
\end{figure}     
In this way, the surviving fibers close to the broken bond are stressed much more than the fibers at a distance, creating a high stress concentration around the broken fiber.
\begin{figure}[ht]
\centering
\includegraphics[width=7cm, keepaspectratio]{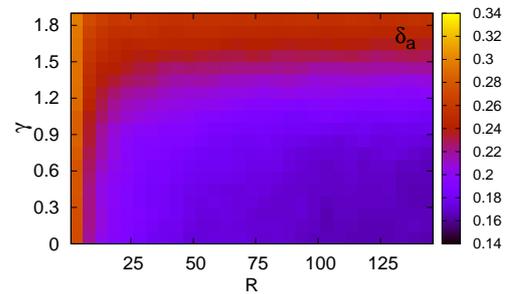}
\caption{(Color online) Study of $\delta_a$ with a continuous variation of $R$ and $\gamma$. Either with decreasing $\gamma$ or increasing $R$ (for $\gamma<\gamma_c$) the model approaches the mean field limit. Above $\gamma_c$ the stress localization sustains at any even for $R>R_c(=L^{2/3})$.}
\label{fig:3d_Plot}
\end{figure}     
According to this redistribution scheme, the load redistributed to $j^{th}$ fiber after the failure of $i^{th}$ fiber is proportional to $1/|i-j|^\gamma$. The stress localization around a broken fiber will be larger for a higher $\gamma$ value. Former studies on fiber bundle model shows that there exists a critical value $\gamma_c$ (1.43 for 1d bundle \cite{Biswas} and 2.0 for 2d bundle \cite{Hidalgo}) below which the model operates in the mean field limit. Here I have presented how the failure abruptness as well as the strength of the bundle responses to disorder and system size for this new stress redistribution scheme. 

Figure \ref{fig:Abruptness_Plwerlaw} shows the variation of $n_c$ with disorder for system size $L=10^4$. The stress release range is kept fixed at a high value ($R=500$) while the $\gamma$ value is continuously changed. With increasing $\gamma$ values the failure process becomes more and more abrupt as well as $\delta_a$ shifts to relatively higher values (similar to what was observed in Fig.\ref{fig:Abruptness}(a) with decreasing $R$ values).
 
\begin{figure}[t]
\centering
\includegraphics[width=8.5cm, keepaspectratio]{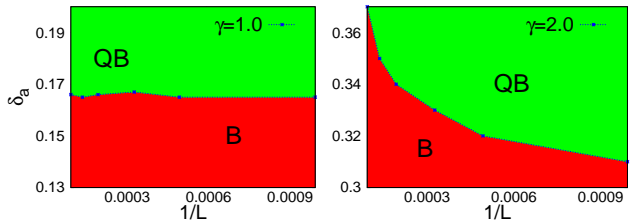}
\caption{(Color online) Variation of $\delta_a$ with system size while $R>R_c$. The plot suggests, we will observe only brittle like failure in the thermodynamic limit for $\gamma>\gamma_c$.}
\label{fig:Dectac_Powerlaw}
\end{figure} 
\begin{figure}[ht]
\centering
\includegraphics[width=6cm, keepaspectratio]{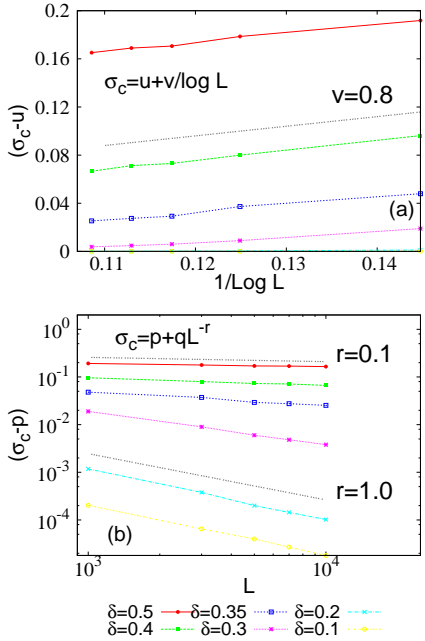}
\caption{(Color online) Variation of critical stress $\sigma_c$ with system size ($L$) at different strength of disorder. The results are fitted with the functions $\Psi_1$ (b) and $\Psi_2$ (a), given by Eq.2.}
\label{fig:System_Size_Powerlaw}
\end{figure} 

Fig.\ref{fig:3d_Plot} describes $\delta_a$, the strength of disorder separating the abrupt and non-abrupt failure, when both $R$ and $\gamma$ is varied. For $\gamma>\gamma_c (\approx1.43)$, the model always acts in the local load sharing limit (dark red) and an increasing $R$ do not lead to the mean field scenario (dark blue). On the other hand for $\gamma<\gamma_c$, the model will either show local stress concentration or operates in the mean field limit depending on the stress release range $R$ (see Ref.\cite{Biswas} for more details). 

I have also revisited the system size effect of $\delta_a$, both below and above $\gamma_c$. $R$ is kept constant at a relatively higher value such that the model is in the mean field limit for $\gamma=0$ (uniform stress redistribution). For $\gamma<\gamma_c$, $\delta_a$ remains constant at its mean field value (=1/6) independent of the system size. Beyond $\gamma_c$, $\delta_a$ gradually increases with increasing system size and we get only brittle like abrupt failure in the thermodynamic limit ($L\rightarrow\infty$), as it is expected and also observed for uniform stress redistribution (see Fig.\ref{fig:Dectac_PhaseDiagram0}). 

Finally I have checked the system size effect of the critical stress at different strength of disorder, similar to what was done in section III(B). To ensure a high stress localization, we have set the $\gamma$ value at 2.0 ($>\gamma_c$). The stress release range $R$ has a value more than $R_c$ so that the stress concentration is introduced in the model through $\gamma$ only. Fig.\ref{fig:System_Size_Powerlaw} shows the variation of $\sigma_c$ with increasing system sizes and for different strength of disorder $\delta$. Under above mentioned condition I have observed the same three regions I, II and III, similar to the case of uniform redistribution (see Fig.\ref{fig:System_Size_Variation1}). At a lower disorder $\sigma_c$ decreases in a scale free manner while in the higher disorder limit the behavior of $\sigma_c$ matches with both scale free as well as inverse logarithmic decrease. The window [$\delta_l,\delta_u$], around which this change in behavior takes place, also remains unchanged for this new stress redistribution scheme.   

%%%%%%%%%%%%%%%%%%%%%%%%%%%%%%%%%%%%%%%%%%%%%%%%%%%%%%%%%%%%%%%%%%%%%%%%%%%%%%%%%

\section{Universality}
To check the universality of the results, I have repeated the study with a power law distribution (with power -1, from $10^{-\beta}$ to $10^{\beta}$) to assign individual threshold value of the fibers. The system size effect on strength and failure abruptness remains the same way, independent of the strength distribution. The results also remain unaltered w.r.t the nature of stress redistribution.  

%%%%%%%%%%%%%%%%%%%%%%%%%%%%%%%%%%%%%%%%%%%%%%%%%%%%%%%%%%%%%%%%%%%%%%%%%%%%%%%%%

\section{Discussion}
The present study shows the effect of system size and disorder on the strength and failure abruptness of a heterogeneous system, namely the fiber bundle model. The strength $\sigma_c$ decreases with system size as $1/L^{\alpha}$ at low disorder and $1/\log{L}$ at moderate disorder value. Also at a low disorder, the failure process is brittle like abrupt, irrespective of the size of the bundle. For high disorder strength ($\delta>\delta^{\ast}$), we get both brittle and quasi-brittle response, depending on the system size. Overall, the study represents different modes of failure at different system size, disorder and stress release range. The finidings are universal w.r.t the distribution of threshold strength values or the nature of stress redistribution.   

%%%%%%%%%%%%%%%%%%%%%%%%%%%%%%%%%%%%%%%%%%%%%%%%%%%%%%%%%%%%%%%%%%%%%%%%%%%%%%%%%

\section{Acknowledgement}
The author thanks Prof. Purusattam Ray, Prof. Bikas Chakrabarty and Dr. Soumyajyoti Biswas for useful discussions at times. Also a spacial thanks goes to Aakriti Saxena Roy for critical reading of the manuscript.

%%%%%%%%%%%%%%%%%%%%%%%%%%%%%%%%%%%%%%%%%%%%%%%%%%%%%%%%%%%%%%%%%%%%%%%%%%%%%%%%%

%%%%%%%%%%%%%%%%%%%%%%%%%%%%%%%%%%%%%%%%%%%%%%%%%%%%%%%%%%%%%%%%%%%%%%%%%%%%%%%%%

\end{document}